\documentclass[acmtocl]{acmtrans2m}

\newdef{definition}[theorem]{Definition}
\newdef{remark}[theorem]{Remark}

\title{Independence of \textit{P} vs. \textit{NP} in regards to oracle relativizations.}
\author{JERRALD MEEK}

\begin{abstract}
This is the third article in a series of four articles dealing with the $P$ vs. \textit{NP} question. The purpose of this work is to demonstrate that the methods used in the first two articles of this series are not affected by oracle relativizations. Furthermore, the solution to the $P$ vs. \textit{NP} problem is actually independent of oracle relativizations.
\end{abstract}


\begin{document}

\begin{bottomstuff}
People who wish to remain anonymous have offered comments and suggestions which have
improved this work. The author wishes to express his appreciation for their assistance.

	\begin{flushright}
		Jerrald Meek Copyright \copyright 2008
	\end{flushright}
\end{bottomstuff}

\maketitle

\section{Introduction.}
Previously in ``$P$ is a proper subset of \textit{NP}'' \protect\cite{meek1} and ``Analysis of the Deterministic Polynomial Time Solvability of the \textit{0-1-Knapsack} Problem'' \protect\cite{meek2} the present author has demonstrated that some \textit{NP-complete} problems are likely to have no deterministic polynomial time solution. In this article the concepts of these previous works will be applied to the relativizations of the $P$ vs. \textit{NP} problem.\par

It has previously been shown by Hartmanis and Hopcroft [1976], that the $P$ vs. \textit{NP} problem is independent of Zermelo-Fraenkel set theory.  However, this same discovery had been shown by Baker [1979], who also demonstrates that if $P =$ \textit{NP} then the solution is incompatible with \textit{ZFC}.\par

	The purpose of this article will be to demonstrate that the oracle relativizations of the $P$ vs. \textit{NP} problem do not preclude any solution to the original problem.  This will be accomplished by constructing examples of five out of the six relativizing oracles from Baker, Gill, and Solovay [1975], it will be demonstrated that inconsistent relativizations do not preclude a proof of $P \neq$ \textit{NP} or $P$ = \textit{NP}. It will then become clear that the $P$ vs. \textit{NP} question is independent of oracle relativizations.

\section{Preliminaries.}
Baker, Gill, and Solovay [1975] have previously shown that there are six types of oracles that can be constructed to examine the $P$ vs. \textit{NP} problem. These include
\begin{enumerate}
	\item Oracle $A$ such that $P^A$ = \textit{NP}$^A$.
	\item Oracle $B$ such that $P^B \neq$ \textit{NP}$^B$.
	\item Oracle $C$ such that \textit{NP}$^C$ is not closed under complementation.
	\item Oracle $D$ such that $P^D \neq$ \textit{NP}$^D$ but \textit{NP}$^D$ is closed under complementation.
	\item Oracle $E$ such that $P^E \neq$ \textit{NP}$^E$ and $P^E =$ \textit{NP}$^E$ $\cap$ \textit{co-NP}$^E$.
	\item Oracle $F$ such that $P^F \subseteq$ [\textit{NP}$^F$ $\cap$ \textit{co-NP}$^F$] $\subseteq$ \textit{NP}.
\end{enumerate}

\newtheorem{thm_opt}{Theorem 4.4 from \textit{P is a proper subset of NP}. \protect\cite{meek1}}[section]
\begin{thm_opt}
\textbf{\emph{P = NP Optimization Theorem.}}\par

The only deterministic optimization of a \textit{NP-complete} problem that could prove \textit{P = NP} would be one that can always solve a \textit{NP-complete} problem by examining no more than a polynomial number of input sets for that problem.
\end{thm_opt}

\begin{definition}
\textbf{\emph{Oracle Machine}}\par

An oracle machine can be deterministic or non deterministic. The difference between an oracle machine and a regular Turing Machine is that an oracle Machine has a yes state, a no state, and a query state. When the oracle machine is placed in the query state after passing a string to the oracle, if that string is an element of the oracle set, then the oracle will place the Turing Machine in the yes state, other wise it will place the Turing Machine in the no state. The processing time required by the oracle can be considered instantaneous.
\end{definition}

\subsection{Encoding Methods.}
In this article, two encoding methods will be used. One encoding method will encode the input sets for a problem and is identical to that used by Baker, Gill, and Solovay [1975]. This will be identified as ``input encoding.''\par

The second method of encoding will be used to identify a partition of the set of all possible input sets for a problem. This method of encoding will be identified as ``partition encoding.''

\subsubsection{Partition Encoding}
The concept of partition encoding is not to encode the input sets for a problem, but an entire partition of the set of all possible input sets. However, for this method to work the partitioning must be identified with a specific problem.\par

The method of partition encoding used in this work will divide partitions by the number of literals in an input set which have a \textit{true} value. If a problem has $n$ literals, then there will be $n + 1$ partitions (one partition has no \textit{true} literals).\par

Notice that the problem $\left[ {a \vee b \vee c} \right]$ has different input sets that result in \textit{true} evaluations from the problem $\left[ {\neg a \vee b \vee c} \right]$ even though the cardinality of these input sets are equivalent.\par

It will then be necessary that the problem which a partition applies to will need to be encoded. This can be accomplished by the following method:
\begin{itemize}
	\item Let $P$ be a \textit{NP-Complete} problem with $k$ literals.
	\item Let $g$ be the G\"{o}del Number calculated from $P$.
	\item Let $\sigma$ be a set of strings representing each of the $k + 1$ partitions for $P$.
	\item Let $1 \leq i \leq k + 1$.
	\item Let $S$ be the set of all possible input sets for $P$.
	\item $T \left( x \right) $ = the number of \textit{true} literals in input set $x$.
\end{itemize}

\noindent then
\[ \forall i : \left( {\sigma_i \equiv \left\langle {i - 1, g} \right\rangle } \right) \]

\noindent and the partition encoding of $S_i$ is represented by:
\[ \sigma_{T \left( {S_i} \right) } \]

\noindent For each partition of $P$ there will be one string consisting of:
\begin{enumerate}
	\item a number in the range of 0 to $k$ representing the number of \textit{true} literals in $S_i$;
	\item and the G\"{o}del Number of $P$.
\end{enumerate}

\section{The Relativizing Oracles.}
Baker, Gill, and Solovay [1975] have shown that the classes of $P$ and \textit{NP} do not relativize in a consistent manner. The inconsistency of the relativizations of $P$ and \textit{NP} has often been used to justify arguments that the $P$ vs. \textit{NP} problem is unusually difficult at best, or maybe even impossible.\par

\begin{quotation}
\noindent By slightly altering the machine model, we can obtain differing answers to the relativized question. This suggests that resolving the original question requires careful analysis of the computational power of machines. It seems unlikely that ordinary diagonalization methods are adequate for producing an example of a language in \textit{NP} but not in $P$; such diagonalizations, we would expect, would apply equally well to the relativized classes, implying a negative answer to all relativized \textit{P =? NP} questions, a contradiction. \protect\cite{baker_gill}
\end{quotation}

Although it is generally accepted that diagonalization is not an effective method for solving the $P$ vs. \textit{NP} problem, it is the purpose of this article to demonstrate that oracle relativizations in no way prevent a solution to the $P$ vs. \textit{NP} problem.\par

It is also worthwhile to mention that the six oracles of Baker, Gill, and Solovay [1975] do not represent an exclusive set of all oracle relativizations. There has been extensive work to find other unusual relativizations of complexity classes in regards to the \textit{P} vs. \textit{NP} question. For example, Bennett and Gill [1981] have found that random oracles can produce the relativization $P^A \neq$ \textit{NP}$^A \neq$ \textit{co-NP}$^A$. However, the scope of this work will be limited to the six oracles from Baker, Gill, and Solovay [1975] with the expectation that what is learned from these six oracles may be applied to any other relativizing oracle.

\subsection{A \textit{P = NP} Oracle.}
The \textit{P = NP Optimization Theorem} requires that no more than a polynomial number of input sets can be examined by a deterministic polynomial time search algorithm. By the same logic it is obvious that a deterministic oracle machine can not request more than a polynomial number of queries in polynomial time.\par

Baker, Gill, and Solovay [1975] define the oracle $A$ such that $A$ = \{$\left\langle {i, x, 0^n} \right\rangle$ : \textit{NP}$_i^A$ accepts $x$ in $< n$ steps\}.\par

This definition produces an oracle such that $P^A =$ \textit{NP}$^A$. However, this definition does not really say anything about how the oracle functions. In this work, the oracle will be created by partition encoding strings, a string will be included in oracle $A$ if there exists at least one element of the partition represented by the string which results in the problem associated to the string evaluating \textit{true}.
\begin{itemize}
	\item Let $L \left( X \right)$ = \{$x$ : there exists $y \in X$ such that $\left| y \right| = \left| x \right|$\} for any oracle $X$.
	\item Let $x$ be the set of all problems in $L \left( X \right)$.
	\item Let $S$ be the set such that $S_i$ is the set of all possible input sets for $x_i$ when $1 \leq i \leq \left| x \right|$.
\end{itemize}

The oracle set is created by the following algorithm.
\begin{enumerate}
	\item Let $i = 1$.
	\item Let $e = 1$.
	\item If $x_i$ evaluates \textit{true} with input $S_{i_e}$, then add the partition encoding of $S_{i_e} \mapsto x_i$ to set $A$, and let $e = \left| S_i \right|$.
	\item If $e < \left| S_i \right|$, then increment $e$ and continue at step 3.
	\item If $e = \left| S_i \right|$ and $i < \left| x \right|$, then increment $i$ and continue at step 2.
	\item If $e = \left| S_i \right|$ and $i = \left| x \right|$, then the set $A$ has been found.
\end{enumerate}

The goal was to produce an example of an oracle set such that $P^A$ = \textit{NP}$^A$. Here, the oracle $A$ is actually such that $P^A$ = \textit{NP}. Such an oracle would also create the condition that $P^A$ = \textit{NP}$^A$. Another method could be used to create an oracle such that $P^X$ = \textit{PSPACE}, which would also satisfy $P^X$ = \textit{NP}$^X$, except that \textit{NP} $\neq$ \textit{NP}$^X$. There are multiple such oracles, however according to Mehlhorn [1973] the set of computable oracles such that $P^A$ = \textit{NP}$^A$ is meager.\par

Notice that in this example the set of all possible input sets will be exponential in size, and therefore some partitions for any problem will be exponential in size. It is then easy to see that the problem of creating the oracle set (even when the oracle is created to only work for one problem) is a member of the \textit{FNP} complexity class (when $P^A$ = \textit{NP}).\par

The process that a Deterministic Oracle Machine with oracle set $A$ uses to solve problem $x_i$ which is in \textit{NP} and has $n$ literals is the algorithm.
\begin{enumerate}
	\item Let $e = 0$.
	\item Encode the partition for problem $x_i$ for input sets having $e$ \textit{true} literals.
	\item Pass the partition encoding from step 2 to the oracle and enter the query state.
	\item If the machine is in the yes state then halt in an accepting state.
	\item If the machine is in the no state and $e < n$ then increment $e$ and continue at step 2.
	\item If the machine is in the no state and $e = n$ then halt in a non-accepting state.
\end{enumerate}

The algorithm of the oracle machine solves any \textit{NP} problem in polynomial time on a Deterministic Oracle Machine. A Non Deterministic Oracle Machine could also use this algorithm to solve the same set of problems in polynomial time. Therefore, $P^A$ = \textit{NP}$^A$.\par

This method is dependant on an oracle set creation algorithm requiring deterministic exponential time for each problem that the oracle is capable of solving. If so desired, the time required to compute the oracle set could be treated as separate to the process of solving the problem.\par

Another option is that the Deterministic Oracle Machine could be networked to both an oracle and a Non Deterministic Turing Machine. In this model the non deterministic machine generates the oracle set which it sends to the oracle. This method will allow the entire process to require polynomial time.\par

A third method is that we simply assume that the oracle set somehow creates itself magically.\par

Regardless of the method used to generate the oracle set, all three methods involve somehow concealing the majority of the work. This does not in any way reduce the complexity of the problem; it only removes the difficulty of the problem by placing the majority of the computational burden upon the process of oracle set creation.\par

In this example the oracle $A$ was created such that $P^A$ = \textit{NP}. It would have been valid to create oracle set $A$ such that $P^A$ = \textit{PSPACE}. However, it can not be expected that such an oracle would be any easier to create.

\subsection{A $P \neq$ \textit{NP} Oracle.}
The method of producing oracle set B as described by Baker, Gill, and Solovay [1975] is similar to the algorithm:
\begin{itemize}
	\item Let $L \left( X \right)$ = \{$x$ : there exists $y \in X$ such that $\left| y \right| = \left| x \right|$\} for any oracle $X$.
	\item Let $x$ be the set of all problems in $L \left( X \right)$.
	\item Let $S$ be the set such that $S_i$ is the set of all possible input sets for $x_i$ when $1 \leq i \leq \left| x \right|$.
	\item Let $B \left( i \right)$ represent the elements of $B$ placed in $B$ prior to stage $i$.
	\item Let $p_i \left( n \right) < \left| S_i \right|$ be a polynomial function of $n$ that represents the number of computations preformed by $P_i^X$ and \textit{NP}$_i^X$ for all oracles $X$. If $p_i \left( n \right)$ computations have been preformed, and no accepting element of $S_i$ has been found, then the algorithm halts in the non-accepting state without examining all elements of $S_i$.\par
\end{itemize}

Create the oracle with the algorithm.
\begin{enumerate}
	\item Let $i = 1$.
	\item If $P_i^B(i)$ rejects $x_i$ then the next unexamined element of $S_i$ is an element of $B$.
	\item If $i < \left| x \right|$ then increment $i$ and continue at step 2.
	\item If $i = \left| x \right|$ then the oracle set $B$ has been found.
\end{enumerate}

The process used by an oracle machine with oracle set $B$ for finding a solution to a problem $x_i$ in \textit{NP} with $n_i$ literals is preformed by the following algorithm.
\begin{enumerate}
	\item Search for an element of $S_i$ which results in $x_i$ evaluating \textit{true}. Terminate the search if the time limit $p_i \left( n \right)$ has been reached.
	\item If a solution to $x_i$ was found then, halt in an accepting state.
	\item If no solution to $x_i$ was found and all elements of $S_i$ were examined in step 1, then halt in a non accepting state.
	\item If no solution to $x_i$ has been found and not all elements of $S_i$ have been examined, then query the oracle $B$ with an unexamined element of $S_i$.
	\item If the machine is in the yes state then halt in an accepting state.
	\item If the machine is in the no state then halt in a non accepting state.
\end{enumerate}

This algorithm will allow for a polynomial time solution on a Non Deterministic Oracle Machine \textit{NP}$^B$ because such a machine will always find a solution in step one for any problem in \textit{NP} if one exists, and halt in step 2 or 3.\par

However, a Deterministic Oracle Machine may not find the solution in step 1. When a deterministic machine queries the oracle the oracle only contains one set of the proper cardinality which may or may not be an accepting input set (because the result of the input set is not a requirement for membership in $B$). The oracle will then place the machine in the yes state if the input set passed to the oracle is equivalent to the one input set of proper cardinality that is an element of $B$. It is then the case that the Deterministic Oracle Machine may terminate in an accepting state when there is no accepting input to the problem. On the other hand if the deterministic oracle machine fails to find a solution in step 1, and the input set passed to the oracle is not an element of $B$, then the machine may terminate in a non-accepting state when an accepting input set may exist. It is then clear that oracle $B$ does not produce a reliable polynomial time solution to the problem $x_i$.\par

It is then true that $P^B \neq$ \textit{NP}$^B$. However this is only because the oracle $B$ is dysfunctional. Oracle $B$ does not actually help a Non Deterministic Oracle Machine to solve a \textit{NP} problem, because the non deterministic machine has the ability to solve the problem without entering the query state. The oracle is then never queried, and the dysfunctionality of the oracle does not interfere with the operation of the Non Deterministic Oracle Machine. However, a Deterministic Oracle Machine requires the assistance of the oracle to overcome the \textit{P = NP Optimization Theorem}. It is then the case that a dysfunctional oracle will prevent a Deterministic Oracle Machine from solving a problem in polynomial time while not affecting the performance of a Non Deterministic Oracle Machine.\par

It then becomes easy to see why $P^A$ = \textit{NP}$^A$ oracle sets are meager \protect\cite{mehlhorn}. The oracle set must be carefully fabricated in order to create such a condition. If great care has not been taken to ensure that the oracle will be of assistance to a deterministic machine, then the oracle will be dysfunctional.

\subsection{An Oracle such that \textit{NP} is not closed under complementation.}

A complexity class is closed under complementation if the complements of all problems in that class are members of the same class. Baker, Gill, and Solovay [1975] have shown that there exists an oracle such that one or more problems in \textit{NP}$^C$ have a complement that is not in \textit{NP}$^C$.
\begin{itemize}
	\item Let $L \left( X \right)$ = \{$x$ : there exists $y \in X$ such that $\left| y \right| = \left| x \right|$\} for any oracle $X$.
	\item Let $x$ be the set of all problems in $L \left( X \right)$.
	\item Let $S$ be the set such that $S_i$ is the set of all possible input sets for $x_i$ when $1 \leq i \leq \left| x \right|$.
	\item Let $C \left( i \right)$ represent the elements of $C$ placed in $C$ prior to stage $i$.
	\item Let $p_i \left( n \right) < \left| S_i \right|$ be a polynomial function of $n$ that represents the maximum number of computations preformed by $P_i^X$ and \textit{NP}$_i^X$ for all oracles $X$. If $p_i \left( n \right)$ computations have been preformed, and no accepting element of $S_i$ has been found, then the algorithm halts in the non-accepting state without examining all elements of $S_i$.
\end{itemize}

Create the oracle with the algorithm.
\begin{enumerate}
	\item Let $i = 1$.
	\item If \textit{NP}$_i^{C \left( i \right)}$ accepts $x_i$ then any one element of $S_i$ that is accepted by $x_i$ is added to the set $C$.
	\item If $i < \left| x \right|$ then increment $i$ and continue at step 2.
	\item If $i = \left| x \right|$ then the oracle set $C$ has been found.
\end{enumerate}

The set $C$ is a set containing exactly one accepting input set for each problem described by $L \left( X \right)$ that has an accepting input set.\par

The process of finding a solution to a problem $x_i$ which is in \textit{NP} with oracle set $C$ is preformed with the following algorithm.
\begin{enumerate}
	\item Pass each element of $S_i$ to the oracle and enter the query state.
	\item If the machine is in the yes state for any query, then halt in an accepting state.
	\item If the machine is in the no state for all queries, then halt in a non accepting state.
\end{enumerate}

For a Non Deterministic Oracle Machine, all elements of $S_i$ can be evaluated simultaneously. However, a Deterministic Oracle Machine must iterate threw them one at a time. It is then easy to see that $P^C \neq$ \textit{NP}$^C$.\par

To create the oracle set $\bar{C}$ the compliment of $C$, use the algorithm.
\begin{enumerate}
	\item Let $i = 1$.
	\item If \textit{NP}$_i^{\bar{C}\left( i \right)}$ rejects $x_i$ then add all elements of $S_i$ to the set $\bar{C}$.
	\item If $i < \left| x \right|$ then increment $i$ and continue at step 2.
	\item If $i = \left| x \right|$ then the oracle set $\bar{C}$ has been found.
\end{enumerate}

With the oracle set $\bar{C}$ a Deterministic Oracle Machine can solve the compliment of any \textit{NP} problem by performing one query. This is because $\bar{C}$ contains all input sets for any problem in \textit{co-NP} that has at least one input set that results in the problem evaluating \textit{true}. If the oracle places the machine in the yes state for any input set then an accepting input set exists for the \textit{co-NP} problem, although the accepting input set may not be represented by the string queried. If the oracle places the machine in the no state then no accepting input set exists for the \textit{co-NP} problem. It is then the case that $P^{\bar{C}} \supset$ \textit{co-NP}.

However, this situation requires that the oracle sets $C$ and $\bar{C}$ are created non deterministically. It is then the case that the work of solving these problems has been done by the creation of the oracle sets, and this allows the Deterministic Oracle Machine to avoid the limitations of the \textit{P = NP Optimization Theorem} when solving \textit{co-NP} problems.

\subsection{A $P \neq$ \textit{NP} oracle such that \textit{NP} is closed under complementation.}
\begin{itemize}
	\item Let $L \left( X \right)$ = \{$x$ : there exists $y \in X$ such that $\left| y \right| = \left| x \right|$\} for any oracle $X$.
	\item Let $x$ be the set of all problems in $L \left( X \right)$.
	\item Let $S$ be the set such that $S_i$ is the set of all possible input sets for $x_i$ when $1 \leq i \leq \left| x \right|$.
	\item Let $D \left( i \right)$ represent the elements of $D$ placed in $D$ prior to stage $i$.
	\item Let $\bar{D} \left( i \right)$ represent the elements of $\bar{D}$ placed in $\bar{D}$ prior to stage $i$.
	\item Let $p_i \left( n \right) < \left| S_i \right|$ be a polynomial function of $n$ that represents the maximum number of computations preformed by $P_i^X$ and \textit{NP}$_i^X$ for all oracles $X$. If $p_i \left( n \right)$ computations have been preformed, and no accepting element of $S_i$ has been found, then the algorithm halts in the non-accepting state without examining all elements of $S_i$.
\end{itemize}

Construct the oracle sets $D$ and $\bar{D}$ with the algorithm.
\begin{enumerate}
	\item Let $i = 1$.
	\item Let $n = 1$.
	\item If $n$ is even, then let $e = 1$.
	\item If $n$ is even and $S_{n_e} \notin \bar{D} \left( i \right)$ then find the prefix $u$ of $S_{n_e}$ with length $\left| S_{n_e} \right| / 2$.
	\item If $n$ is even and $S_{n_e} \notin \bar{D} \left( i \right)$ and $u$ is an input set for $x_{ \left| u \right| }$ and \textit{NP}$_i^{D \left( n \right)}$ rejects $x_{\left| u \right|}$, then $S_{n_e} \in D$.
	\item If $n$ is even and $S_{n_e} \notin \bar{D} \left( i \right)$ and $e < \left| S_n \right|$ then increment $e$ and continue at step 4.
	\item If $n$ is even and $S_{n_e} \notin \bar{D} \left( i \right)$ and $e = \left| S_n \right|$ then break to step 10.
	\item If $n$ is odd and all elements of $\bar{D} \left( n \right)$ have cardinality less than $n$, and $p_i \left( n \right) < 2^{\left( {n - 1} \right) / 2}$, then add to $\bar{D}$ all strings queried by $P_i^{D \left( n \right)}$ for problem $x_n$. If $P_i^{D \left( n \right)}$ rejects $x_n$, then add to $D$ the next element of $S_n$ not queried.
	\item If $n$ is odd, then increment $i$.
	\item If $n < \left| x \right|$, then increment $n$ and continue at step 3.
	\item If $n = \left| x \right|$, then the sets $D$ and $\bar{D}$ have been found.
\end{enumerate}

In steps 4 and 5, elements are added to $D$ if their prefix is rejected. Notice that if $f \left( x \right)$ has no input set resulting in a \textit{true} evaluation, then $f \left( x \right) \wedge g \left( x \right)$ has no input set resulting in a \textit{true} evaluation. It is then the case that in step 4 and 5, elements are only added to $D$ if they represent a problem that has no accepting input set.\par

In step 8, an input set is added to $D$ if $P_i^{D \left( n \right)}$ rejects the problem which that input set belongs to. However, on the first iteration, $D \left( n \right)$ will be an empty set. An empty oracle set is equivalent to having no oracle, and so the polynomial time algorithm will not check all possible input sets for the first problem examined. For the third problem examined, $D \left( n \right)$ may contain an input set if it belongs to a problem that always evaluates \textit{false}. It is then easy to see that step 8 adds elements to $D$ that may or may not cause a problem to evaluate \textit{true}.\par

Because the elements of $D$ do not have any consistent representation, it then follows that the set $D$ will result in a dysfunctional oracle. A deterministic machine requires a functional oracle set to be able to solve all \textit{NP} problems, but a non deterministic machine does not rely upon the oracle set when solving \textit{NP} problems. It then follows that $P^D \neq$ \textit{NP}$^D$.\par

In step 8, the requirement that all elements of $\bar{D} \left( n \right)$ must have cardinality less than $n$ will always evaluate \textit{true}. The statement $p_i \left( n \right) < 2^{\left( {n - 1} \right) / 2}$ will evaluate \textit{true} for the first values of $n$ up to some limit. For problems less than this limit, a polynomial number of elements will be added to $\bar{D}$. The elements will be added regardless of weather or not they result in a \textit{true} evaluation of the \textit{co-NP} problem associated to them. It is then clear to see that the set $\bar{D}$ will also result in a dysfunctional oracle, and $P^{\bar{D}}$ will not produce reliable results for any \textit{co-NP} problem.

\subsection{A $P \neq$ \textit{NP} oracle such that $P$ = [\textit{NP} $\cap$ \textit{co-NP}].}
Baker, Gill, and Solovay [1975] construct the oracle set $E$ by adding elements to the oracle set $A$ which was previously constructed such that $P^A$ = \textit{NP}$^A$. The additional elements in $E$ will result in $P^E \neq$ \textit{NP}$^E$. The trick is to arrange the additional elements such that if a problem is in \textit{NP}, and the compliment of the problem is also in \textit{NP}, then both of these problems are solvable in polynomial time by the oracle $P^E$.\par

In the description of creating oracle $E$, it is allowed to start with ``any oracle $A$ such that $P^A$ = \textit{NP}$^A$.'' Here, the oracle $A$ that was previously constructed in this article will be used. However, remember that oracle $A$ did not contain strings representing input sets, but instead partition encoding was used. It will then be the case that the same encoding specification will need to be used for the additional elements of $E$.
\begin{itemize}
	\item Let $L \left( X \right)$ = \{$x$ : there exists $y \in X$ such that $\left| y \right| = \left| x \right|$\} for any oracle $X$.
	\item Let $x$ be the set of all problems in $L \left( X \right)$.
	\item Let $S$ be the set such that $S_i$ is the set of all possible input sets for $x_i$ when $1 \leq i \leq \left| x \right|$.
	\item Let $e \left( n \right)$ be a function such that $e \left( 0 \right) = 0$ and $e \left( x \right) = 2^{2e \left( {x - 1} \right)} \leftarrow x > 1$.
	\item Let $E \left( 0 \right) = A$ and let $E \left( n \right)$ be the set of elements placed in $E$ prior to stage $n$.
	\item Let $r$ be the set of all possible values for $\left\langle {j, k} \right\rangle \leftarrow j \in N, k \in N, j \neq k$.
	\item Let $r_{j \left( i \right)}$ denote the $j$ element of set $r_i$, and $r_{k \left( i \right)}$ denote the $k$ element of $r_i$.
	\item Let $p_i \left( n \right) < \left| S_i \right|$ be a polynomial function of $n$ that represents the maximum number of computations preformed by $P_i^X$ and \textit{NP}$_i^X$ for all oracles $X$. If $p_i \left( n \right)$ computations have been preformed, and no accepting element of $S_i$ has been found, then the algorithm halts in the non-accepting state without examining all elements of $S_i$.
\end{itemize}

Create the oracle set $E$ with the algorithm.
\begin{enumerate}
	\item Let $n = 1$.
	\item Let $i = 1$.
	\item Let $l \in S_n$.
	\item If $e \left( {n-1} \right) < \left( {log_2 \left| l \right|} \right) \leq e \left( n \right) \leq max \left\{ {p_{r_{j \left( i \right)}}} \left( l \right), p_{r_{k \left( i \right)}} \left( l \right) \right\} < e \left( {n + 1} \right)$ and neither \textit{NP}$_{r_{j \left( i \right)}}^{E \left( n \right)}$ or \textit{NP}$_{r_{k \left( i \right)}}^{E \left( n \right)}$. accepts $x_n$, then $\left\langle {j, k} \right\rangle$ is unsatisfied.
	\item If $p_i \left( {e \left( n \right)} \right) \geq 2^{e \left( n \right) }$, then $\left\langle {i, i} \right\rangle$ is unsatisfied.
	\item If $\left\langle {j, k} \right\rangle$ was not determined unsatisfied in step 4, and $\left\langle {i, i} \right\rangle$ was not determined unsatisfied in step 5, then increment $i$.
	\item If $\left\langle {j, k} \right\rangle$ was not determined unsatisfied in step 4, and $\left\langle {i, i} \right\rangle$ was determined unsatisfied in step 5, and $P_i^{E \left( n \right) }$ rejects $x_n$, then add to $E$ the encoding of $x_n$ with the next unevaluated element of $S_n$.
	\item If $\left\langle {j, k} \right\rangle$ was not determined unsatisfied in step 4, and $\left\langle {i, i} \right\rangle$ was determined unsatisfied in step 5, then increment $i$.
	\item If $i < \left| N \right|$ then increment $n$ and continue at step 3.
	\item If $i = \left| N \right|$ then the set $E$ has been found.
\end{enumerate}

Notice that when $\left\langle {i, i} \right\rangle$ is unsatisfied, then $P_i^{E \left( n \right)}$ will be run on $x_n$. The set $E \left( n \right)$ will only contain elements relevant to $x_n$ if it was inherited from set $A$. It is then the case that $P_i^{E \left( n \right) }$ will only reject $x_n$ when $x_n$ has no accepting input set. In step 7, if $\left\langle {i, i} \right\rangle$ is unsatisfied, and $P_i^{E \left( n \right) }$ rejects $x_n$, then the input set encoding for $x_n$ will always be an encoding that represents a partition containing no accepting input set. It will then be the case that $P^E$ will accept $x_n$ when a Non Deterministic Oracle Machine would halt in a non accepting state without querying the oracle. Therefore, $P^E \neq$ \textit{NP}$^E$ because the oracle $E$ is dysfunctional when handling some problems.\par

Let $\kappa$ be the set of all problems in \textit{NP} which have complements in \textit{NP}. When $x_n \in \kappa$, then $\left\langle {j, k} \right\rangle$ will be satisfied. When $\left\langle {j, k} \right\rangle$ is satisfied then no strings applicable to $x_n$ will be added to $E$. It is then the case that the subset of $E$ that applies to all elements of $\kappa$ will be identical to the subset of $A$ that applies to all elements of $\kappa$. So $P^E$ = \textit{NP}$^E$ when the problem being evaluated is an element of $\kappa$.

\subsection{An oracle set such that $P^F \subseteq$ [\textit{NP}$^F$ $\cap$ \textit{co-NP}$^F$] $\subseteq$ \textit{NP}.}

Baker, Gill and Solovay [1975] do not lay out the specific method for creating oracle $F$. However they do provide an outline which includes creating two languages $L_1 \left( F \right)$ and $L_2 \left( F \right)$ such that $L \left( F \right) = \left[ {L_1 \left( F \right) \cup L_2 \left( F \right)} \right]$. Two oracle sets must also be created by two different methods, and $F$ is the union of these two oracle sets.\par

However, notice that if $P^F \subseteq$ [\textit{NP}$^F$ $\cap$ \textit{co-NP}$^F$], and [\textit{NP}$^F$ $\cap$ \textit{co-NP}$^F$] $\subseteq$ \textit{NP}, then $P^F \subseteq$ of \textit{NP}.\par

It has already been demonstrated that an oracle set can be created such that $P^A$ = \textit{NP}. In other words the set of problems solvable by a Deterministic Oracle Machine with oracle set $A$ in polynomial time is equivalent to the set of problems solvable by a Non Deterministic Turing Machine in polynomial time.\par

It is then the case that the only new condition of oracle set $F$ is that [\textit{NP}$^F$ $\cap$ \textit{co-NP}$^F$] contains all problems in \textit{NP}. This is the same thing as saying that \textit{NP} is strictly contained within \textit{co-NP}.\par

Earlier, it was shown that an oracle set $C$ could be created such that \textit{NP}$^C$ is not closed under complementation. The reason that oracle set $C$ allowed this condition was because it resulted in all problems in \textit{co-NP} being solvable in polynomial time by $P^{\bar{C}}$, while problems in \textit{NP} where not solvable in polynomial time by $P^C$.\par

It is then easy to see that an oracle $F$ could be created such that all problems in \textit{NP} are solvable in polynomial time by $P^F$, and all problems in \textit{co-NP} are solvable in polynomial time by $P^{\bar{F}}$. Under such an oracle the complexity classes \textit{NP}, and \textit{co-NP} are both contained by $P$.\par

If \textit{co-NP} $\subset P^{\bar{F}}$, and \textit{NP} $\subseteq P^F$, then $P^F \subseteq$ [\textit{NP} $\cap$ \textit{co-NP}]. All problems in both \textit{NP} and \textit{co-NP} have polynomial time solutions on a Deterministic Oracle Machine with oracle set $F$. It then follows that all problems in \textit{NP} and \textit{co-NP} have polynomial time solutions on a Non Deterministic Oracle Machine with oracle set $F$.

\section{Oracle relativizations and P vs. NP solution methods.}
Here, the relation of oracle relativizations to the $P$ vs. \textit{NP} question will be examined by creating an analog of the $P$ vs. \textit{NP} question. Two complexity classes will be created, $P_{\Lambda}$ and \textit{NP}$_{\Lambda}$; the question will be asked, is $P_{\Lambda}$ = \textit{NP}$_{\Lambda}$?\par

To create the analog complexity classes, one problem that is obviously in $P$ will be excluded from $P_{\Lambda}$. Any problem in $P$ will due, here the problem used will be the \textit{set-sum} problem. The \textit{set-sum} problem will be defined as:
\begin{itemize}
	\item Let $S$ be a set of real numbers.
	\item Let $r$ be the number of elements in $S$.
	\item Let $M$ be a real number.
\end{itemize}

	\[ \sum_{i=1}^r S_i = M \]

Given the value of $S$ and $M$, determine if the equality evaluates \textit{true}.\par
\vspace{1ex}

Notice that the \textit{set-sum} problem is similar to the \textit{0-1-Knapsack} problem, and could be solved by finding the sum of all subsets of $S$, but only evaluate \textit{true} if the one subset of $S$ that contains all elements of $S$ sums to $M$. In the analog complexity model, it will be assumed that this is the only known method of solving \textit{set-sum}.\par

An analog complexity model will be created which ignores the existence of some \textit{NP} problems (such as \textit{NP-complete} problems). The analog complexity classes will be defined as:
\begin{itemize}
	\item Let \textit{NP}$_{\Lambda}$ contain all problems known to be in $P$ assuming $P \neq NP$.
	\item Let $P_{\Lambda}$ contain all problems in \textit{NP}$_{\Lambda}$ except the \textit{set-sum} problem.
\end{itemize}

In the analog computational model \textit{NP}$_{\Lambda}$ contains all problems known to be solvable in non deterministic polynomial time, while $P_{\Lambda}$ contains all problems known to be solvable in deterministic polynomial time. In the analog question $P_{\Lambda}$ = \textit{NP}$_{\Lambda}$ assume that no method has yet been found to prove that \textit{set-sum} is a member of $P_{\Lambda}$, yet \textit{set-sum} has not been conclusively excluded from $P_{\Lambda}$.\par
\vspace{1ex}

Now the following questions may be asked.
\begin{itemize}
	\item Does there exist an oracle such that $P_{\Lambda}^A$ = \textit{NP}$_{\Lambda}^A$?
	\item Does there exist an oracle such that $P_{\Lambda}^B \neq$ \textit{NP}$_{\Lambda}^B$?
	\item Does there exist an oracle such that \textit{NP}$_{\Lambda}^C$ is not closed under complementation?
	\item Does there exist an oracle such that $P_{\Lambda}^D \neq$ \textit{NP}$_{\Lambda}^D$ but \textit{NP}$_{\Lambda}^D$ is closed under complementation?
	\item Does there exist an oracle such that $P_{\Lambda}^F \subseteq$ [ \textit{NP}$_{\Lambda}^F \cap$ \textit{co-NP}$_{\Lambda}^F$ ] $\subseteq$ \textit{NP}$_{\Lambda}$?
\end{itemize}

It should be easy to see that the answer to all of these questions is yes. If inconsistent relativizations are an indication of the difficulty of the $P_{\Lambda}$ vs. \textit{NP}$_{\Lambda}$ problem, then the problem of proving the \textit{set-sum} problem solvable in deterministic polynomial time must be at least as difficult as proving that \textit{NP-complete} problems are solvable in deterministic polynomial time.\par

Is the set of computable oracles such that $P_{\Lambda}^X$ = \textit{NP}$_{\Lambda}^X$ meager? If so, then does that indicate a greater likelihood that $P_{\Lambda} \neq$ \textit{NP}$_{\Lambda}$?\par

With this analog it happens to be the case that $P_{\Lambda}$ = \textit{NP}$_{\Lambda}$. If the $P$ vs. \textit{NP} question is dependant on oracle relativizations, then the analog $P_{\Lambda}$ vs. \textit{NP}$_{\Lambda}$ question should indicate that \textit{P = NP}. If $P$ vs. \textit{NP} is dependant on oracle relativizations, then it should be expected that \textit{P = NP} could be proven by finding the relation between the polynomial time solvability of \textit{set-sum} and the relativizations of $P_{\Lambda}$ and \textit{NP}$_{\Lambda}$. If such a relation exists, then it should provide an indication of how to prove \textit{P = NP}. However, if no such relation exists, then there must exist no dependancy between oracle relativizations and the solution to the $P$ vs. \textit{NP} question.\par

Obviously, the proof that $P_{\Lambda}$ = \textit{NP}$_{\Lambda}$ can be accomplished by eliminating the evaluation of all subsets of $S$ for the \textit{set-sum} problem and only evaluating the one relevant subset. It is known that the only relevant subset is the subset equivalent to $S$. This is known because the original definition of the problem says it is so. Is there some way that the oracle relativizations also identify the one relevant subset of $S$? If so then the relativizations should identify a limited partition of relevant subsets for the \textit{Knapsack} problem.\par

It is then easy to see that oracle relativizations only tell us that the complexity of any given problem can be hidden by oracle set creation. Essentially, using an oracle to solve a \textit{NP-complete} problem in deterministic polynomial time is the reverse operation of taking a problem from the $P$ complexity class and pretending that it has no deterministic polynomial time solution. When applied to a \textit{NP-complete} problem, which may actually have no deterministic polynomial time solution, an oracle such that $P^A$ = \textit{NP}$^A$ allows us to pretend that \textit{NP-complete} problems have deterministic polynomial time solutions.

\section{Conclusion.}
In this article all six oracles form Baker, Gill and Solovay [1975] have been examined. As a result, the following things can be said about the oracles.\par
\begin{itemize}
	\item Oracle set $A$ such that $P^A$ = \textit{NP}$^A$ is a functional oracle which allows a Deterministic Oracle Machine to solve any \textit{NP} problem with a polynomial number of queries.
	\item Oracle set $B$ such that $P^B \neq$ \textit{NP}$^B$ is a dysfunctional oracle which fails to allow a Deterministic Oracle Machine to solve all \textit{NP}$^B$ problems with a polynomial number of queries.
	\item Oracle set $C$ such that \textit{NP}$^C$ is not closed under complementation is a dysfunctional oracle such that a Deterministic Oracle Machine can not solve all \textit{NP} problems with a polynomial number of queries, although a Deterministic Oracle Machine could solve any \textit{co-NP} problem with one query.
	\item Oracle set $D$ such that $P^D \neq$ \textit{NP}$^D$ but \textit{NP}$^D$ is closed under complementation is a dysfunctional oracle with a dysfunctional complement. It is then the case that a Deterministic Oracle Machine can not solve any \textit{NP} problem or any \textit{co-NP} problem in polynomial time.
	\item Oracle set $E$ such that $P^E \neq$ \textit{NP}$^E$ and $P^E$ = [\textit{NP}$^E \cap$ \textit{co-NP}$^E$] is a dysfunctional oracle that allows a Deterministic Oracle Machine to solve a \textit{NP} problem in polynomial time when the compliment of the problem is also in \textit{NP}.
	\item Oracle set $F$ such that $P^F \subseteq$ [\textit{NP}$^F$ $\cap$ \textit{co-NP}$^F$] $\subseteq$ \textit{NP} is a functional oracle which allows a Deterministic Oracle Machine to solve all problems in \textit{NP} and all problems in \textit{co-NP} in polynomial time.
\end{itemize}

If \textit{P = NP} then all problems in \textit{NP} will be solvable in deterministic polynomial time with or without an oracle set. It would then be the case that a Deterministic Oracle Machine would only be dependant on an oracle set to solve a problem in \textit{NP} or \textit{co-NP} if the machine is not using an optimal algorithm. It is possible to create an algorithm for any \textit{NP} problem such that a solution would require exponential time on a Non Deterministic Turing Machine. This however does not disqualify the problem's membership in \textit{NP}. Likewise, just because a \textit{NP} problem has a deterministic exponential time solution, this does not disqualify that problem's membership in $P$. Therefore, the oracle relativizations are no indicator of $P$ vs. \textit{NP}.\par

If $P \neq$ \textit{NP} then any problem not ordinarily in $P$ will be solvable in polynomial time by a Deterministic Oracle Machine only when the oracle set is functional for that problem. In this case if $P \neq$ \textit{NP} then the Baker, Gill, and Solovay model works exactly as would be expected.\par

The meagerness of computable $P^X$ = \textit{NP}$^X$ oracle sets also seems not to indicate anything about the $P$ vs. \textit{NP} problem. That is, we should expect a functional oracle for any purpose to be meager. All of these observations lead to the conclusion that the $P$ vs. \textit{NP} question is independent of oracle relativizations.

\section{Version history.}
The author wishes to encourage further feedback which may improve, strengthen, or perhaps disprove the content of this article. For that reason the author does not publish the names of any specific people who may have suggested, commented, or criticized the article in such a way that resulted in a revision, unless premission has been granted to do so.

\noindent \textbf{arXiv Current Version}\newline
3Sep08 Submitted to arXiv.
\begin{itemize}
	\item Revision of partition encoding method.
	\item Refrance to \protect\cite{nagel} was added.
\end{itemize}

\noindent \textbf{arXiv Version 5}\newline
22Aug08 Submitted to arXiv.
\begin{itemize}
	\item Minor addition.
\end{itemize}

\noindent \textbf{arXiv Version 4}\newline
21May08 Submitted to arXiv.
\begin{itemize}
	\item Minor error correction.
\end{itemize}

\noindent \textbf{arXiv Version 3}\newline
20May08 Submitted to arXiv.
\begin{itemize}
	\item Clarifications were made for some statements.
	\item The \textit{Oracle relativizations and P vs. NP solution methods} section was added.
	\item Refrance to \protect\cite{bennett} and \protect\cite{mehlhorn} were added.
\end{itemize}

\noindent \textbf{arXiv Version 2}\newline
16May08 The article was withdrawn due to misleading statements and incomplete research.

\noindent \textbf{arXiv Version 1}\newline
14May08 Submitted to arXiv.

\begin{received}
Received xx/2008; revised xx/2008; accepted xx/2008
\end{received}
\end{document}